\documentclass[letter,twocolumn]{jpsj3}
\usepackage{txfonts}
\usepackage{braket}

\title{Basis of spinors expressed by differential forms and calculating its norm}

\author{Daisuke A. Takahashi$^{1,2}$\thanks{takahashi@phys.chuo-u.ac.jp,\ daisuke.takahashi@keio.jp}}
\inst{$^1$Department of Physics, Chuo University, 1-13-27 Kasuga, Bunkyo-ku, Tokyo 112-8551, Japan \\
$^2$Research and Education Center for Natural Sciences, Keio University, Hiyoshi 4-1-1, Yokohama, Kanagawa 223-8521, Japan} 

\abst{
The basis of spinors in three-dimensional Euclidean space is expressed by differential forms. Its expression is found from the spectral decomposition of the modified Hamiltonian describing Weyl semimetals where the wavenumber parameters are replaced by the differential forms. The generalization of the definitions of differential forms including not only fractional powers but also any algebraic functions is justified by algebraic extension of the symmetric Fock space isomorphic to multivariable polynomial ring. We furthermore define the inner product for these fractional-order differential forms written in two ways; the formal power series and the multiple integral. While the norms of the powers $ \mathrm{d}s^{2\nu} $ are reduced to the variant of Dyson's integral and have the value $2\nu+1$, possibly related to the dimension of irreducible representation, the norm of two-component spinor is given by the integral
\[ I=\frac{2}{\pi}\int_{x^2+y^2\ge 1 \& 0\le x \le 1 \& 0 \le y \le 1} \frac{xy\mathrm{d}x\mathrm{d}y}{\sqrt{(1-x)(1-y)(x^2+y^2-1)}}=\frac{4\sqrt{2}}{\pi}\int_0^1 \frac{(1+\!\!\sqrt{z})}{(1+z)^3}\big[2E(z)-(1-z)K(z)\big]\mathrm{d}z\simeq 1.774, \]
 where $ K(z) $ and $ E(z) $ are the complete elliptic integrals of the first and second kind, respectively. The product can also be generalized to include one parameter $ p $ analogous to those in $L^p$ space, reducing to the original one when $ p=2 $. The norm with $ p=\infty $ is considered as an example and the result for $ \nu=-\frac{1}{2} $ is written by the Watson-Iwata integral.
  We also discuss the ambiguity of the definition of the spinor under coordinate rotation originating from Berry's phase, and point out that if we heuristically set $ \mathrm{d}s=0 $, the ambiguity disappears, though its implication remains unclear. 
}

\begin{document}
\maketitle

In traditional textbooks of the theory of relativity, the vectors and tensors are defined by their linear transformation rules under the coordinate transformation, which, in today's standard theory of manifold\cite{Nakahara}, are understood as an expansion coefficient using the basis of the (co)tangent space naturally associated with the choice of the coordinate. This knowledge tells us a genuine mathematical identification of the vectors which we write $ \vec{e}_x, \vec{e}_y, \vec{e}_z $ in the undergraduate lecture of electromagnetism and accept intuitively as an ``arrow'' without precise definition; that is, their actual nature is the differential operators $ \partial_x,\partial_y,\partial_z $ for the tangent space and the differential forms  $ \mathrm{d}x, \mathrm{d}y, \mathrm{d}z $ for the cotangent space.\\
\indent How about spinors? In fact, the representation theory of SO(3) found in the textbooks of quantum mechanics also gives the ``basisless'' expression of spinors, which is analogous to the above-mentioned traditional treatment of vectors; if one gets the two-dimensional representation of rotation matrices, then the spin-$\frac{1}{2}$ particle is defined by the transformation rule under the multiplication of these $2\times 2$ representation matrices. 
We thus reach the question that what is an actual entity of this geometrical object possessing these two components as expansion coefficients.
One may expect that this question will be soon solved if we know how spinors have been formulated in curved spacetime. However, the vielbein formalism\cite{Nakahara} avoids this problem in a tricky way, where one temporarily prepares a local orthonormal system by diagonalizing the metric tensor point by point, and basisless spinors formulated once in a flat space as they are are placed on it. 
This method makes it possible to discuss physical phenomena concerning spinors in curved space, but the problem of identification of spinor becomes obscured. One might even think that this basisless formulation is the essential nature of spinor and considering such question is meaningless. However, we in reality encounter the spinor whose basis is explicitly written by the square root of the differential form in conformal field theory\cite{EguchiSugawara}, where, in the general coordinate transformation $ x=x(z) $, the fermion field transforms like $ \psi(z)=\left( \frac{\mathrm{d} x}{\mathrm{d} z} \right)^{1/2}\psi(x) $, which may be formally written as $ \psi(z)(\mathrm{d}z)^{1/2}=\psi(x)(\mathrm{d}x)^{1/2} $, indicating that it is a ``rank-$\frac{1}{2}$ covariant tensor''. It simply suggests that the composite of two spin-$\frac{1}{2}$ is a spin-1, i.e., a vector. In one dimension, the dimensions of the (co)tangent space are also 1, so there remains no degree of freedom in making the ``square root of a basis''. However, if dimension gets larger, what kind of combination of basis $ \mathrm{d}x^1,\mathrm{d}x^2,\dots,\mathrm{d}x^n $ is a suitable definition becomes nontrivial. Here we provide a possible solution in three dimension.

The starting observation is that the covariant vector $ A=\sum_{i=1}^3 A_i\mathrm{d}x^i $ can be expressed as a trace of the Pauli matrices:
\begin{align}
	A_i\mathrm{d}x^i = \tfrac{1}{2}\operatorname{Tr} \left[ (A_i\sigma^i)  (\sigma_j \mathrm{d}x^j)   \right], \label{eq:paulitrace}
\end{align}
where $ \sigma_i=\sigma^i $ are Pauli matrices and henceforth summation symbols for repeated indices are sometimes omitted. We also use the notation $ (\mathrm{d}x^1,\mathrm{d}x^2,\mathrm{d}x^3)=(\mathrm{d}x,\mathrm{d}y,\mathrm{d}z) $ depending on the situation. Let us now consider the column and row spinor $ \psi=\left( \begin{smallmatrix} \psi_\uparrow \\ \psi_\downarrow \end{smallmatrix} \right) $ and $ \phi=\left( \begin{smallmatrix}\phi_\uparrow & \phi_\downarrow \end{smallmatrix} \right) $. Their product $ \psi\phi $ is a $ 2\times 2 $ matrix possessing the irreducible decomposition into spin-0 and spin-1, and the latter, following Eq.~(\ref{eq:paulitrace}), is given by
\begin{align}
	&\phi (\sigma_j \mathrm{d}x^j) \psi  =  \operatorname{Tr}\left[ \psi \phi (\sigma_j \mathrm{d}x^j)  \right]=(\phi_\uparrow\psi_\downarrow+\phi_\downarrow\psi_\uparrow)\mathrm{d}x \notag \\
	&\qquad\qquad\quad -\mathrm{i}(\phi_\uparrow\psi_\downarrow-\phi_\downarrow\psi_\uparrow)\mathrm{d}y+(\phi_\uparrow\psi_\uparrow-\phi_\downarrow\psi_\downarrow)\mathrm{d}z.   \label{eq:spinorprod}
\end{align}
Our present aim is to find the basis of spinor written by differential forms. Looking at Eq.~(\ref{eq:spinorprod}) carefully, we expect that
a desired thing will be obtained by ``cutting'' the form $ (\sigma_j\mathrm{d}x^j) $ into two parts.
That is, writing the spectral decomposition
\begin{align}
	\sigma_j \mathrm{d}x^j = \mathrm{d}s \left( v_0 v_0^\dagger-v_1 v_1^\dagger \right),\quad \mathrm{d}s \coloneqq \!\sqrt{\mathrm{d}x^2+\mathrm{d}y^2+\mathrm{d}z^2}, \label{eq:spinordecomp}
\end{align}
where the decomposition is performed by regarding $ \mathrm{d}x^j $'s as ordinary real numbers (this will be justified later), 
then Eq.~(\ref{eq:spinorprod}) is given by $ \phi (\sigma_j \mathrm{d}x^j) \psi = \mathrm{d}s\left[ \phi v_0 v_0^\dagger \psi - \phi v_1 v_1^\dagger \psi  \right] $, and the column and row spinors with basis expressed by differential forms would be constructed as
\begin{align}
	\sqrt{\mathrm{d}s}\, v_j^\dagger \psi &= \psi_{\uparrow} \sqrt{\mathrm{d}s}\,v_{j,\uparrow}^*+ \psi_{\downarrow} \sqrt{\mathrm{d}s}\,v_{j,\downarrow}^*, \label{eq:spinorbasis1} \\
	\sqrt{\mathrm{d}s}\, \phi v_j &= \phi_{\uparrow} \sqrt{\mathrm{d}s}\,v_{j,\uparrow}+ \phi_{\downarrow} \sqrt{\mathrm{d}s}\, v_{j,\downarrow},
\end{align}
where we write $ v_j=\left( \begin{smallmatrix} v_{j,\uparrow} \\ v_{j,\downarrow} \end{smallmatrix} \right) $. 
Thus, the ket vectors $ \ket{\uparrow} $ and $ \ket{\downarrow} $, which we learned in quantum mechanics, would be identified as --- or at least proportional to --- the above $ \sqrt{\mathrm{d}s}\,v_{j,\uparrow}^* $ and $ \sqrt{\mathrm{d}s}\, v_{j,\downarrow}^* $. 

Let us investigate the decomposition (\ref{eq:spinordecomp}) in more detail. In fact, if the form $ \mathrm{d}x^j $ are replaced by an ordinary number $ x^j $, it reduces to the diagonalization of the Weyl Hamiltonian
\begin{align}
	H(\boldsymbol{x})=\sum_{i=1}^3 x^i \sigma_i = \begin{pmatrix} x^3 & x^1-\mathrm{i}x^2 \\ x^1+\mathrm{i}x^2 & -x^3 \end{pmatrix}, \label{eq:weyl}
\end{align}
which, if the parameter $ \boldsymbol{x}=(x^1,x^2,x^3) $ is interpreted as a Bloch wavenumber $\boldsymbol{k}$, is equivalent to the model of Weyl semimetals\cite{AndoTopo}. It is also a minimal model where the adiabatic change of the system parameter yields a nonvanishing phase factor for quantum states, i.e., Berry's phase [Ref.~\citen{Nakahara}, Chap. 10], thus exhibiting topological phenomena.

In order to justify the diagonalization of the Weyl Hamiltonian where differential forms are substituted $ H(\mathrm{d}\boldsymbol{x}) $ instead of ordinary numbers, we must define algebraic operations for $ \mathrm{d}x^j $'s including divisions, fractional powers, etc. It is achieved by considering the symmetric algebra, which is isomorphic to multivariable polynomial ring [Ref. \citen{SatakeEng}, Chap. V]. This algebra is also equivalent to the Fock space of bosonic many-body systems. Then, the division is easily defined by that of polynomial, and moreover, making an algebraic extension by adding zeros of polynomials, we obtain the algebraic-function field. By this extension, we can use the square root of differential forms unambiguously.\\
\indent Next, we define the inner product for this space. For a while, we consider an abstract $d$-dimensional vector space spanned by a basis $ e_1,\dots,e_d $. We also write the basis of its dual space $ f_1,\dots,f_d $ satisfying $ \braket{e_i,f_j}=\delta_{ij} $. Then, the inner product between rank-$ r $ tensors is given by $ \braket{ e_{i_1}\otimes \dots\otimes e_{i_r} , f_{j_1}\otimes \dots \otimes f_{j_r} }=\delta_{i_1j_1}\dots \delta_{i_r j_r} $. The correspondence between polynomials and symmetric tensors is as follows\cite{SatakeEng}. Regarding $ e_1,\dots,e_d $ as indeterminates, the monomial $ e_1^{n_1}e_2^{n_2}\dots e_d^{n_d},\ n_1,n_2,\dots, n_d \in\mathbb{Z}_{\ge0} $, corresponds to  $ e_1^{n_1}e_2^{n_2}\dots e_d^{n_d} = \mathcal{S} (e_1^{\otimes n_1}\otimes e_2^{\otimes n_2}\otimes \dots \otimes e_d^{\otimes n_d}) $, where we write $ e^{\otimes n} \coloneqq e\otimes\dots \otimes e \, (\text{$n$ times}) $ 
 and $ \mathcal{S} $ represents a projection operator to the symmetric tensor space, satisfying $ \mathcal{S}^2=\mathcal{S} $ and acting as $ \mathcal{S}(e_{i_1}\otimes \dots\otimes e_{i_r}) = \frac{1}{r!}\sum_{\sigma\in\mathfrak{S}_r}e_{i_{\sigma(1)}}\otimes \dots\otimes e_{i_{\sigma(r)}} $. 
The same correspondence also holds for $ f_1^{m_1}f_2^{m_2}\dots f_d^{m_d},\ m_1,m_2,\dots,m_d\in\mathbb{Z}_{\ge0} $. The inner product of these two is then calculated as 
\begin{align}
	\braket{ e_1^{n_1}\dots e_d^{n_d}, f_1^{m_1}\dots f_d^{m_d} } = \delta_{n_1m_1}\dots \delta_{n_dm_d} \binom{\sum_{i=1}^d n_i}{n_1,\dots,n_d}^{-1}, \label{eq:syminpr}
\end{align}
where $ \binom{\sum_{i=1}^d n_i}{n_1,\dots,n_d}=\frac{(\sum_{i=1}^d n_i)!}{n_1!\dots n_d!} $ is a multinomial coefficient.\\
\indent Henceforth we only consider $ d=3 $ though keeping general $ d $ is not hard. Let $ A(e_1,e_2,e_3) $ and $ B(f_1,f_2,f_3) $ be  $ n $-th order homogeneous functions, i.e., the functions s.t. $ e_3^nA(\frac{e_1}{e_3},\frac{e_2}{e_3},1)=A(e_1,e_2,e_3) $ and $ f_3^nB(\frac{f_1}{f_3},\frac{f_2}{f_3},1)=B(f_1,f_2,f_3) $. First let us assume $ n\in\mathbb{Z}_{\ge0} $ and $ A $ and $ B $ are polynomial functions. Then, writing $ \xi_i=e_i/e_3,\ \eta_i=f_i/f_3,\ i=1,2 $, they have expansion $ A=e_3^n \sum_{n_1,n_2\in\mathbb{Z}_{\ge0}} a_{n_1,n_2} \xi_1^{n_1}\xi_2^{n_2} $ and $ B=f_3^n \sum_{m_1,m_2in\mathbb{Z}_{\ge0}} b_{m_1,m_2}\eta_1^{m_1}\eta_2^{m_2} $ and the inner product is given by
\begin{align}
	\braket{A,B}= \sum_{n_1,n_2\in\mathbb{Z}_{\ge0}} a_{n_1,n_2}b_{n_1,n_2}\binom{n}{n_1,n_2,n-n_1-n_2}^{-1}. \label{eq:syminpr2}
\end{align}
If $ A $ and $ B $ are polynomials the range of this summation is finite: $ \sum_{0 \le n_1+n_2 \le n} $. On the other hand, when $ n\not\in\mathbb{Z}_{\ge0} $ and/or $ A $ and $ B $ include rational and/or fractional-power functions, their Taylor series generally becomes infinite. Even in this case, the inner product defined by Eq.~(\ref{eq:syminpr2}) makes sense. However, if we actually calculate this formal power series, the result often diverges. So, we want another definition equivalent to this one but has a larger convergence region. We can achieve it if we find a linear operator giving the mapping
\begin{align}
	e_1^{n_1}e_2^{n_2}e_3^{n_3} f_1^{m_1}f_2^{m_2}f_3^{m_3} \to \delta_{n_1m_1}\delta_{n_2m_2}\delta_{n_3m_3}\binom{n}{n_1,n_2,n_3}^{-1}.
\end{align}
It is indeed realized by setting $ e_i=\sqrt{\rho_i}\mathrm{e}^{\mathrm{i}u_i},\ f_i=\sqrt{\rho_i}\mathrm{e}^{-\mathrm{i}u_i} $ and applying the integral operator $ \frac{1}{n!}\prod_{i=1}^3 \int_0^\infty \mathrm{d}\rho_i \mathrm{e}^{-\rho_i} \int_0^{2\pi}\frac{\mathrm{d}u_i}{2\pi} $. Thus, the integral expression of the inner product is given by
\begin{align}
	&\braket{A,B} = \frac{1}{n!}\prod_{i=1}^3 \int_0^\infty \!\!\mathrm{d}\rho_i \mathrm{e}^{-\rho_i}\! \int_{-\pi}^{\,\pi}\frac{\mathrm{d}u_i}{2\pi}  \notag \\
	\!\!\!\!&A(\!\!\sqrt{\rho_1}\mathrm{e}^{\mathrm{i}u_1}\!,\!\!\sqrt{\rho_2}\mathrm{e}^{\mathrm{i}u_2}\!,\!\!\sqrt{\rho_3}\mathrm{e}^{\mathrm{i}u_3})B(\!\!\sqrt{\rho_1}\mathrm{e}^{-\mathrm{i}u_1}\!,\!\!\sqrt{\rho_2}\mathrm{e}^{-\mathrm{i}u_2}\!,\!\!\sqrt{\rho_3}\mathrm{e}^{-\mathrm{i}u_3}),\! \label{eq:syminpr3}
\end{align}
which often gives a finite value even when the series (\ref{eq:syminpr2}) diverges. Replacing the quantum inner product by a classical multiple integral reminds us of path integral. The two of the six-fold integral in Eq.~(\ref{eq:syminpr3}) can be soon performed; first, by shifting $ u_1 \to u_1+u_3  $ and $ u_2 \to u_2+u_3 $, the integrand becomes $ u_3 $-independent, and next, introducing the polar coordinate $ (\rho_1,\rho_2,\rho_3)=\gamma^2(\hat{n}_1^2,\hat{n}_2^2,\hat{n}_3^2) $ with $ (\hat{n}_1,\hat{n}_2,\hat{n}_3)=(\sin\beta\cos\alpha,\sin\beta\sin\alpha,\cos\beta) $, the integration w.r.t. $ \gamma $ reduces to $ \Gamma $ function. Therefore, the integral which we must actually calculate is four-fold. We also note that if the integrand in Eq.~(\ref{eq:syminpr3}) only contains the even powers of $ \mathrm{e}^{\pm\mathrm{i}u_i} $'s, the replacement $ \mathrm{e}^{\pm\mathrm{i}u_i} \to \mathrm{e}^{\pm\mathrm{i}u_i/2} $ does not change the result and simplifies the integrand. The examples considered below all belong to this case and this replacement is made unless otherwise noted. \\
\indent Note that this algebra can be equipped with another inner product. Let $ p>0 $ be a rational number. If the above-described algebraic extension starts from the vector space spanned by $ e_1^{p/2},\dots,e_d^{p/2} $ instead of $ e_1,\dots,e_d $, the final algebraic field contains the same elements but the product (\ref{eq:syminpr}) is modified by
\begin{align}
	\Braket{ e_1^{n_1}\dots e_d^{n_d}, f_1^{n_1}\dots f_d^{n_d} }_p \coloneqq \binom{\frac{2}{p}\sum_{i=1}^d n_i}{\frac{2}{p}n_1,\dots,\frac{2}{p}n_d}^{-1}. \label{eq:psyminpr2-1}
\end{align}
The original one (\ref{eq:syminpr}) corresponds to $ p=2 $. The product itself can be defined even for irrational $ p $. 
In particular, the limit $ p\to\infty $ is
\begin{align}
	\Braket{ e_1^{n_1}\dots e_d^{n_d}, f_1^{n_1}\dots f_d^{n_d} }_\infty =1, \label{eq:psyminpr2-2}
\end{align} 
and the integral expression realizing this inner product is 
\begin{align}
	\!\!\!\!\braket{A,B}_{\infty} \!=\! \prod_{i=1}^3 \!\int_{-\pi}^{\,\pi}\!\frac{\mathrm{d}u_i}{2\pi} \,A(\mathrm{e}^{\mathrm{i}u_1}\!,\mathrm{e}^{\mathrm{i}u_2}\!,\mathrm{e}^{\mathrm{i}u_3})B(\mathrm{e}^{-\mathrm{i}u_1}\!,\mathrm{e}^{-\mathrm{i}u_2}\!,\mathrm{e}^{-\mathrm{i}u_3}).\!\! \label{eq:syminpr3inf}
\end{align}
The product $ \braket{\cdot,\cdot}_p $ is invariant under the orthogonal transformation $ (e_i')^{p/2}=R_{ij}e_j^{p/2},\ (f_i')^{p/2}=R_{ij}f_j^{p/2} $ with $ R $ an orthogonal matrix, which is partially similar to those in $L^p$ space. Hence, the product with $ p=2 $ has prime importance in physical application.  \\
\indent Let us finish abstract algebra and go back to differential forms. Replacing $ e_i \to \mathrm{d}x^i $ and $ f_i \to \frac{\partial }{\partial x^i} $, the expressions shown above can be used. As a basic example, let us consider the norm of $ \mathrm{d}s^{2\nu}=(\mathrm{d}x^2+\mathrm{d}y^2+\mathrm{d}z^2)^\nu $. The dual element is given by $ \nabla^{2\nu}=(\partial_x^2+\partial_y^2+\partial_z^2)^\nu $. One can prove
\begin{align}
	\braket{\mathrm{d}s^{2\nu},\nabla^{2\nu}}=2\nu+1,\quad (\nu>-\tfrac{3}{2}). \label{eq:dsnorm1}
\end{align}
The derivation of Eq.~(\ref{eq:dsnorm1}), expanding the integrand, reduces to the formula
\begin{align}
	\!\mbox{\footnotesize $ {\displaystyle \prod_{i=1}^3\!\int_{-\pi}^{\,\pi}\!\frac{\mathrm{d}u_i}{2\pi}} \sin^{2l_1}\!\frac{u_2-u_3}{2}\sin^{2l_2}\!\frac{u_3-u_1}{2}\sin^{2l_3}\!\frac{u_1-u_2}{2}\!=\!\frac{(l_1+l_2+l_3)!(l_1-\frac{1}{2})!(l_2-\frac{1}{2})!(l_3-\frac{1}{2})!}{\pi^{3/2}(l_1+l_2)!(l_2+l_3)!(l_3+l_1)!}\!,$} \label{eq:varDy}
\end{align}
which is a variant of Dyson's integral\cite{10.1063/1.1703773} but eludes the proof by Good\cite{10.1063/1.1665339} (see also Ref.~\citen{AAR}, \S8.8); to calculate Eq.~(\ref{eq:varDy}) we need the constant term of $ \prod_{j\ne l}(1-\frac{z_j}{z_l})^{a_{j,l}} $ with $ a_{j,l}=a_{l,j} $ but their theorem only treats the case $ a_{j,l}=a_j $.  \\
\indent The $ \infty $-norm $ \braket{\mathrm{d}s^{2\nu},\nabla^{2\nu}}_\infty $ has finite value when $ \nu>-1 $. In particular, the case $ \nu=-\frac{1}{2} $ can be explicitly evaluated as
\begin{align}
	\braket{\tfrac{1}{\sqrt{\mathrm{d}s^2}},\tfrac{1}{\sqrt{\nabla^2}}}_\infty=\tfrac{2\sqrt{3}}{\pi^2}K(\sin^2\!\tfrac{\pi}{12})^2=\tfrac{1}{6\pi^2}B(\tfrac{1}{2},\tfrac{1}{6})^2\simeq 0.896, \label{eq:dsnorm4}
\end{align}
where  $ K(z) $ is the complete elliptic integral of the first kind and $ B(x,y) $ is the beta function. 
This integral reduces to those studied by Watson and Iwata\cite{10.1093/qmath/os-10.1.266,oai:teapot.lib.ocha.ac.jp:00034938,BGMWZ}. 
The power series given below suggests that all integrals with other half-odd integers $ \nu\ge-\frac{1}{2} $ reduce to elliptic integrals.\\ 
\indent It is worth observing what happens if the formal power series (\ref{eq:syminpr2}) is used. We consider general $ p $-norm so we replace the multinomial coefficient by (\ref{eq:psyminpr2-1}) or (\ref{eq:psyminpr2-2}). By binomial expansion $ \mathrm{d}s^{2\nu}=\mathrm{d}z^{2\nu}\sum_{l=0}^\infty\binom{\nu}{l}\sum_{r=0}^l\binom{l}{r}\xi_1^{2r}\xi_2^{2l-2r} $, the power series of the norm is obtained but diverges. So, we temporarily introduce an auxiliary variable $ x^l $ in the summand, yielding
\begin{align}
	\textstyle\braket{\mathrm{d}s^{2\nu},\nabla^{2\nu}}_p \mathop{\text{``$=$''}}\, {\displaystyle\sum_{l=0}^\infty \sum_{r=0}^l} \binom{\nu}{l}^2\binom{l}{r}^2 \binom{\frac{4}{p}\nu}{\frac{4}{p}(\nu-l),\,\frac{4}{p}(l-r),\,\frac{4}{p}r}^{-1}x^l,
\end{align}
where $ \mathop{\text{``$=$''}} $ means that the limit $ x\to1 $ is taken at the last stage. This summation can be carried out for $ p=2 $ and $\infty $: 
\begin{align}
	\braket{\mathrm{d}s^{2\nu},\nabla^{2\nu}} &\mathop{\text{``$=$''}} F^2_1\left( \begin{smallmatrix} 1,-\nu \\ \frac{1}{2}-\nu \end{smallmatrix};x \right), \label{eq:dsfps1} \\
	\braket{\mathrm{d}s^{2\nu},\nabla^{2\nu}}_\infty &\mathop{\text{``$=$''}} F^3_2\left( \begin{smallmatrix} \frac{1}{2},-\nu,-\nu \\ 1,1 \end{smallmatrix};4x \right), \label{eq:dsfps2}
\end{align}
where $ F^p_q\left( \begin{smallmatrix} a_1,\dots,a_p \\ b_1,\dots,b_q  \end{smallmatrix};z \right)\coloneqq \sum_{l=0}^\infty \frac{(a_1)_l\dots(a_p)_l}{(1)_l(b_1)_l\dots(b_q)_l}z^l $ is the generalized hypergeometric function. If $ \nu\in\mathbb{Z}_{\ge0} $, both (\ref{eq:dsfps1}) and (\ref{eq:dsfps2}) return a correct answer since they reduce to a polynomial. 
The expansion of  Eq.~(\ref{eq:dsfps1}), $ \frac{\sqrt{\pi}\,\Gamma(\frac{1}{2}-\nu)}{\Gamma(-\nu)}\frac{1}{\sqrt{1-x}}\!+\!(2\nu\!+\!1)\!+\!O(\!\!\sqrt{1-x}) $, reproduces Eq.~(\ref{eq:dsnorm1}) if the first divergent term is excluded. 
Equation (\ref{eq:dsfps2}) substituted $ x=1 $ has a complex value, which is inconsistent with the integral (\ref{eq:syminpr3inf}) whose result is real and positive, implying something wrong in analytic continuation. More concrete information can be obtained for half-odd integer $ \nu $'s; using Clausen's formula and the contiguous relations [Ref.~\citen{ErdelyiHTF1}, Chap. IV], Eq. (\ref{eq:dsfps2}) reduces to the elliptic integral. Furthermore, we can check that its real part coincides with the integral (\ref{eq:syminpr3inf}) numerically. Indeed, $\text{Eq.~(\ref{eq:dsfps2})}_{\nu=-1/2} =F^3_2\left( \begin{smallmatrix} \frac{1}{2},\frac{1}{2},\frac{1}{2} \\ 1,1 \end{smallmatrix};4x \right)=\tfrac{4}{\pi^2}K\Big(\frac{1-\sqrt{1-4x}}{2}\Big)^2 $, with $ x=1 $ and Landen's transformation, reduces to $ \frac{4}{\pi^2}K(\mathrm{e}^{-\mathrm{i}\pi/3})^2=\frac{4}{\pi^2}K(\sin^2\!\frac{\pi}{12})\mathrm{e}^{-\mathrm{i}\pi/6}, $ whose real part is (\ref{eq:dsnorm4}). \\
\indent Now we again consider writing the basis of spinors by differential forms. Introducing the spherical-coordinate-like parametrization $ (\mathrm{d}x,\mathrm{d}y,\mathrm{d}z)=\!\!\sqrt{\mathrm{d}s^2}(\sin\theta\cos\varphi,\sin\theta\sin\varphi,\cos\theta) $, which is now unambiguously defined by algebraic extension with $ \theta,\varphi $ being not ordinary numbers, the diagonalization of Eq.~(\ref{eq:weyl}) with differential forms substituted is performed in the same way as that of ordinary numbers:
\begin{align}
	H(\mathrm{d}\boldsymbol{x})&= \sqrt{\mathrm{d}s^2}V(\mathrm{d}\boldsymbol{x}) \sigma_3  V(\mathrm{d}\boldsymbol{x})^\dagger,  \label{eq:weyl20} \\
	V(\mathrm{d}\boldsymbol{x})&\coloneqq \mathrm{e}^{\frac{-\mathrm{i}}{2}\varphi\sigma_3}\mathrm{e}^{\frac{-\mathrm{i}}{2}\theta\sigma_2}. \label{eq:weyl21}
\end{align}
Using the column and row spinors $ \psi $ and $ \phi $, we can make an SU(2)-invariant $ \phi H(\mathrm{d}\boldsymbol{x})\psi = \mathrm{d}s \phi V(\mathrm{d}\boldsymbol{x}) \sigma_3 V(\mathrm{d}\boldsymbol{x})^\dagger \psi $. From this expression, we introduce the covariant column spinor $ (\mathrm{d}s^2)^{\frac{1}{4}} V(\mathrm{d}\boldsymbol{x})^\dagger \psi $ and the covariant row spinor $ (\mathrm{d}s^2)^{\frac{1}{4}} \phi V(\mathrm{d}\boldsymbol{x}) $ as a spinor whose basis is written by differential forms. We also simply call them the covariant ket and bra.
If we write $ V=(v_0,v_1), $ we easily find $ v_j(\theta\pm\pi)=\pm (-1)^j v_{1-j}(\theta) $.  The above definition includes an array of two possible bases $ v_0 $ or $ v_1 $; either of two can be used, 
but we should carefully consider Berry's phase discussed later. 
In the spherical coordinate $(r,\theta,\varphi)$ of normal numbers, if we formally allow negative $ r $, the parametrization has the redundancy of replacement $ (r,\theta,\varphi)\to (-r,\theta+\pi,\varphi) $, reflecting the 1:2 correspondence of SO(3) and SU(2). By this replacement the exchange of the eigenvalues and eigenvectors $ r \leftrightarrow -r $ and $ v_0 \leftrightarrow v_1 $ occurs. Since we normally restrict the range of  $ \theta $ to $ [0,\pi] $, if $ v_0 $ is the ``nearside'' basis, $ v_1 $ is the ``farside'' one. \\
\indent The contravariant ket and bra are also introduced in the same way: $ (\nabla^2)^{1/4} V(\nabla)^\dagger \psi $ and $ (\nabla^2)^{1/4} \phi V(\nabla) $. 
The norm density is then obtained by the product of contravariant bra and covariant ket $  \psi^\dagger \Braket{ (\nabla^2)^{\frac{1}{4}} V(\nabla), V(\mathrm{d}\boldsymbol{x})^\dagger (\mathrm{d}s^2)^{\frac{1}{4}} } \psi $.  If we write  $ \Braket{ (\nabla^2)^{\frac{1}{4}} V(\nabla),\ V(\mathrm{d}\boldsymbol{x})^\dagger (\mathrm{d}s^2)^{\frac{1}{4}} }=I_0 \boldsymbol{1}_2+\sum_{i=1}^3 I_i \sigma_i $, we can check that $ I_1,I_2,I_3 $ all vanish because the power series only includes the oscillating terms. $ \sqrt{\mathrm{d}s}\,v_0 $ and $ \sqrt{\mathrm{d}s}\,v_1 $ are orthogonal in this sense. The only non-vanishing integral $ I_0 $ is
 \begin{align}
 	I_0=\!\frac{1}{(\frac{1}{2})!}\prod_{j=1}^3\int_0^\infty \!\mathrm{d}\rho_i\mathrm{e}^{-\rho_i}\!\!\!\int_{-\pi}^\pi\!\frac{\mathrm{d}u_i}{2\pi} \sqrt{N_3\left( \tfrac{1}{2}+\tfrac{\rho_3+N_2}{2N_3} \right)\left( \tfrac{1}{2}+\tfrac{\rho_1+\rho_2}{2N_2} \right)},
 \end{align}
 where $ N_3 =\left|\rho_1\mathrm{e}^{\mathrm{i}u_1}+\rho_2\mathrm{e}^{\mathrm{i}u_2}+\rho_3\mathrm{e}^{\mathrm{i}u_3}\right|,\ N_2 =\left|\rho_1\mathrm{e}^{\mathrm{i}u_1}+\rho_2\mathrm{e}^{\mathrm{i}u_2}\right| $. 
 A little calculation reveals the following expressions:
  \begin{align}
  	&I_0=\frac{2}{\pi}\int_{D} \frac{xy \mathrm{d}x\mathrm{d}y}{\sqrt{(1-x)(1-y)(x^2+y^2-1)}} \\
  	&=\frac{4\sqrt{2}}{\pi}\int_0^1 \frac{(1+\!\!\sqrt{z})}{(1+z)^3}\big[2E(z)-(1-z)K(z)\big]\mathrm{d}z\simeq 1.774,
  \end{align}
  where $ D $ is $ x^2+y^2\ge 1 \& 0\le x \le 1 \& 0 \le y \le 1 $, and $ K(z) $ and $ E(z) $ are the complete elliptic integral of the first and second kind. The closed form of this integral is unknown.

\indent Finally, we discuss the ambiguity originating from Berry's phase under coordinate rotation. This problem is more serious than the same problem for the ordinary number (\ref{eq:weyl}); in the latter, we only observe the hysteresis as an actual physical phenomenon under actual change of parameters, but in the former, what we do is nothing but a change of frame, so the physical output must be unchanged. Let $ L_1,L_2, $ and $ L_3 $ be generators of rotation with respect to $ x,y, $ and $ z $ axis in three dimension, e.g.,  $ L_3=\sigma_2 \oplus 0 $, and write the element of SO(3) by the Euler angle $ R=\mathrm{e}^{-\mathrm{i}L_3 \alpha }\mathrm{e}^{-\mathrm{i}L_2\beta}\mathrm{e}^{-\mathrm{i}L_3\gamma}. $ The corresponding element in SU(2) is  $ U=\mathrm{e}^{-\mathrm{i}\frac{\sigma_3}{2} \alpha }\mathrm{e}^{-\mathrm{i}\frac{\sigma_2}{2}\beta}\mathrm{e}^{-\mathrm{i}\frac{\sigma_3}{2}\gamma}. $ By the coordinate rotation $ \boldsymbol{x}' = R\boldsymbol{x} $, the corresponding differential form also changes as  $ \mathrm{d}\boldsymbol{x}' = R \mathrm{d}\boldsymbol{x} $. We can soon check
\begin{align}
	H(R\mathrm{d}\boldsymbol{x})=UH(\mathrm{d}\boldsymbol{x})U^\dagger. \label{eq:weyl22}
\end{align}
On the other hand, if the diagonalization is performed from the beginning in the $ \boldsymbol{x}' $-coordinate, we get
\begin{align}
	H(R\mathrm{d}\boldsymbol{x})=\sqrt{\mathrm{d}s^2} V(R\mathrm{d}\boldsymbol{x})\sigma_3 V(R\mathrm{d}\boldsymbol{x})^\dagger. \label{eq:weyl23}
\end{align}
Since Eq. (\ref{eq:weyl22}) with (\ref{eq:weyl20}) must be equal to (\ref{eq:weyl23}), and since the eigenvector with nondegenerate eigenvalue is unique up to an overall factor, there exists a diagonal matrix $ D $ such that
\begin{align}
	V(R\mathrm{d}\boldsymbol{x})D=UV(\mathrm{d}\boldsymbol{x}).
\end{align}
Let $ \theta' $ and $ \varphi' $ be angle parameters for the $ \boldsymbol{x}' $-coordinate. Then, if we write $ D=\mathrm{e}^{\mathrm{i}\xi\sigma_3}, $ the expression for Berry's phase $ \xi $ is found as
\begin{align}
	\mathrm{e}^{2\mathrm{i}\xi}=\frac{\cos\theta\cos(\varphi+\gamma)\sin\beta+\cos\beta\sin\theta-\mathrm{i}\sin\beta\sin(\varphi+\gamma)}{\sin\theta'}.
\end{align}
Since $ \cos\theta'=\cos\beta\cos\theta-\cos(\varphi+\gamma)\sin\beta\sin\theta $, this is indeed a unit complex number ($|\mathrm{e}^{\mathrm{i}\xi}|=1$), but not identically 1. However, if we rewrite this expression using $ \cos\theta=\frac{\mathrm{d}z}{\mathrm{d}s},\ \mathrm{e}^{\pm\mathrm{i}\varphi}=\frac{\mathrm{d}x\pm\mathrm{i}\mathrm{d}y}{\sqrt{\mathrm{d}s^2-\mathrm{d}z^2}}, $ and \textit{heuristically} set $ \mathrm{d}s \to0 $, we obtain 
\begin{align}
	\mathrm{e}^{2\mathrm{i}\xi}\!=\! \mbox{$\small\displaystyle \frac{\sin\beta(\frac{(\cos\gamma\mathrm{d}x-\sin\gamma\mathrm{d}y)\mathrm{d}z-\mathrm{i}\mathrm{d}y\mathrm{d}s}{\sqrt{\mathrm{d}s^2-\mathrm{d}z^2}})+\cos\beta\sqrt{\mathrm{d}s^2-\mathrm{d}z^2}}{\sqrt{\mathrm{d}s^2-(\cos\beta\mathrm{d}z-\cos\gamma\sin\beta\mathrm{d}x+\sin\gamma\sin\beta\mathrm{d}y)^2}} \overset{\mathrm{d}s\to0}{\longrightarrow} 1.$}
\end{align}
Thus, if we would have some justification for this replacement, non-uniqueness caused by Berry's phase would be eliminated.\\
\indent The above heuristic replacement also works for vectors obtained by the composition of two spinors. The composition of ket and bra $ (\!\!\sqrt{\mathrm{d}s} v_i^\dagger \psi) (\!\!\sqrt{\mathrm{d}s}\phi v_j) $ with $ i,j=0,1 $ can be treated by matrix $ V^\dagger \psi \phi V \mathrm{d}s $ all at once; introducing the notation $ \psi\phi= \frac{1}{2}(n\boldsymbol{1}_2+\sum_{i=1}^3\Psi_i \sigma_i) $, we obtain
 $ V^\dagger \psi\phi V\mathrm{d}s= \tfrac{1}{2}n \left(\begin{smallmatrix} \mathrm{d}s & \\ &\mathrm{d}s  \end{smallmatrix}\right)+\tfrac{1}{2}\Psi_3 \left(\begin{smallmatrix} \mathrm{d}z & -\sqrt{\mathrm{d}s^2-\mathrm{d}z^2} \\ -\sqrt{\mathrm{d}s^2-\mathrm{d}z^2} & -\mathrm{d}z \end{smallmatrix}\right)+\tfrac{1}{2}\Psi_1 \left(\begin{smallmatrix} \mathrm{d}x & \frac{\mathrm{d}x\mathrm{d}z+\mathrm{i}\mathrm{d}s\mathrm{d}y}{\sqrt{\mathrm{d}s^2-\mathrm{d}z^2}} \\ \frac{\mathrm{d}x\mathrm{d}z-\mathrm{i}\mathrm{d}s\mathrm{d}y}{\sqrt{\mathrm{d}s^2-\mathrm{d}z^2}} & -\mathrm{d}x \end{smallmatrix}\right)+\tfrac{1}{2}\Psi_2 \left(\begin{smallmatrix} \mathrm{d}y & \frac{\mathrm{d}y\mathrm{d}z-\mathrm{i}\mathrm{d}s\mathrm{d}x}{\sqrt{\mathrm{d}s^2-\mathrm{d}z^2}} \\ \frac{\mathrm{d}y\mathrm{d}z+\mathrm{i}\mathrm{d}s\mathrm{d}x}{\sqrt{\mathrm{d}s^2-\mathrm{d}z^2}} & -\mathrm{d}y \end{smallmatrix}\right). $ 
While the diagonal components, which are the gauge-invariant combination $ v_j^\dagger  $ and $ v_j $ and Berry's phase cancels out, successfully yield the spin-0 and spin-1 components, the off-diagonal components, the composition of $ v_j^\dagger $ and $ v_{1-j} $, depend on Berry's phase and do not provide an ordinary scalar and vector. However, if we set $ \mathrm{d}s \to 0 $, we get $ V^\dagger \psi\phi V\mathrm{d}s \to \frac{1}{2}\sum_{j=1}^3 \Psi_j\mathrm{d}x^j \left( \begin{smallmatrix}1 \\ \mathrm{i}\end{smallmatrix} \right)\left( \begin{smallmatrix}1 & \mathrm{i}\end{smallmatrix} \right) $, thus all four components present vectors. The composition of two kets can also be considered. Let $ \psi,\chi $ be both column spinors. The composition of the gauge-invariant combination $ v_0^\dagger $ and $ v_1^\dagger $ successfully generates the scalar and vector: $ (\!\!\sqrt{\mathrm{d}s}v_0^\dagger\psi)(\!\!\sqrt{\mathrm{d}s}v_1^\dagger \chi)=\tfrac{1}{\sqrt{2}}\Big( \Phi_{1,1}\tfrac{-(\mathrm{d}x+\mathrm{i}\,\mathrm{d}y)}{\sqrt{2}}+\Phi_{1,0}\mathrm{d}z+\Phi_{1,-1}\tfrac{\mathrm{d}x-\mathrm{i}\mathrm{d}y}{\sqrt{2}}+\Phi_{0,0}\mathrm{d}s \Big), $ 
where $ (\Phi_{1,1},\Phi_{1,0},\Phi_{1,-1},\Phi_{0,0})=(\psi_\uparrow\chi_\uparrow, \frac{\psi_\uparrow\chi_\downarrow+\psi_\downarrow\chi_\uparrow}{\sqrt{2}},\psi_\downarrow\chi_\downarrow,\frac{\psi_\uparrow\chi_\downarrow-\psi_\downarrow\chi_\uparrow}{\sqrt{2}}) $ is the spin-triplet and singlet components. On the other hand, the composition of the gauge-dependent pair, two $ v_j^\dagger $'s, yields
 $ (\!\!\sqrt{\mathrm{d}s}v_j^\dagger\psi)(\!\!\sqrt{\mathrm{d}s}v_j^\dagger \chi)=\tfrac{1}{\sqrt{2}}\Big[ \tfrac{(\mathrm{d}x+\mathrm{i}\mathrm{d}y)(\mathrm{d}s+(-1)^j\mathrm{d}z)}{\sqrt{2}\sqrt{\mathrm{d}s^2-\mathrm{d}z^2}}\Phi_{1,1}+(-1)^j{\scriptstyle\sqrt{\mathrm{d}s^2-\mathrm{d}z^2}}\Phi_{1,0}+\tfrac{(\mathrm{d}x-\mathrm{i}\mathrm{d}y)(\mathrm{d}s-(-1)^j\mathrm{d}z)}{\sqrt{2}\sqrt{\mathrm{d}s^2-\mathrm{d}z^2}}\Phi_{1,-1} \Big] $, 
which does not reduce to the covariant vector, but if we set $ \mathrm{d}s \to 0 $, we obtain a vector $ \frac{\mathrm{i}(-1)^j}{\sqrt{2}}( \Phi_{1,1}\tfrac{-(\mathrm{d}x+\mathrm{i}\,\mathrm{d}y)}{\sqrt{2}}+\Phi_{1,0}\mathrm{d}z+\Phi_{1,-1}\tfrac{\mathrm{d}x-\mathrm{i}\mathrm{d}y}{\sqrt{2}}) $. \\
\indent A concluding remark is on the implication of the matrix $ \sigma_j\mathrm{d}x^j $ in Eq.~(\ref{eq:spinordecomp}). In physics, vectors and matrices appear as an equivalent representation of bases of geometrical objects. However, $ \sigma_j\mathrm{d}x^j $ already includes the basis in its matrix components. Therefore, these Pauli matrices are not an array of expansion coefficients of physical quantities but a purely algebraic notion expressing symmetry and structure of physical systems. In fact, in tensor networks and matrix product states\cite{Matsueda}, matrices are already heavily used in this way, where they consider matrices whose component itself is a ket vector; for example, the ground state of the AKLT model is known to reduce the matrix product state. Thus the spinor decomposition presented in this manuscript might provide a novel dissection tool of quantum states in these fields.\\ 
\indent The detailed calculations of the integrals will be submitted elsewhere. Several closely related concepts would be found in other fields including, e.g.,  spin geometry and twistor theory, and identifying these connections is left for future work.
\begin{acknowledgment}
The work was supported by JSPS Grant JP19H05821.
\end{acknowledgment}


\begin{thebibliography}{9}
\bibitem{Nakahara} M.~Nakahara, {\em Geometry, Topology, and Physics},  IOP Publishing, Bristol, England, 2nd edition, 2003.
\bibitem{EguchiSugawara} T. Eguchi and Y. Sugawara, {\em Conformal Field Theory}, Iwanami, 2015. [written in Japanese.]
\bibitem{AndoTopo} Y. Ando, {\em Introduction to topological insulators}, Kodansha, 2014. [written in Japanese.]
\bibitem{SatakeEng} I.~Satake, {\em Linear Algebra}, Marcel Dekker, New York, 1975. [original Japanese version published from Shokabo, 1974.]
\bibitem{10.1063/1.1703773} F.~J. Dyson, {\em Statistical Theory of the Energy Levels of Complex Systems. I}, { J. Math. Phys.}, 3(1):140--156, 01 1962.
\bibitem{10.1063/1.1665339} I.~J. Good, {\em Short Proof of a Conjecture by Dyson}, {J. Math. Phys.}, 11(6):1884--1884, 06 1970.
\bibitem{AAR}G.~Andrews, R.~Askey, and R.~Roy, {\em Special Functions (Encyclopedia of Mathematics and its Applications, Series Number 71)}, Cambridge University Press, Cambridge, 1999.
\bibitem{10.1093/qmath/os-10.1.266} G.~N. Watson, {\em Three triple integrals}, {The Quarterly Journal of Mathematics}, os-10(1):266--276, 01 1939.
\bibitem{oai:teapot.lib.ocha.ac.jp:00034938} G.~Iwata, {\em Evaluation of the Watson Integral of a Face-centered Lattice}, { Natural Science Report, Ochanomizu University}, 20:13--18, 1969.
\bibitem{BGMWZ} J.~M. Borwein, M.~L. Glasser, R.~C. McPhedran, J.~G. Wan, and I.~J. Zucker, {\em Lattice Sums Then and Now (Encyclopedia of Mathematics and its Applications, Series Number 150)}, Cambridge University Press, Cambridge, 2013.
\bibitem{ErdelyiHTF1} A.~Erd{\'e}lyi{\,\,}(ed.),  {\em Higher Transcendental Functions - Volume I - Based, in part, on notes left by Harry Bateman}, McGraw-Hill, New York, 1953.
\bibitem{Matsueda} H.~Matsueda, {\em Entanglement and geometry in quantum systems --- Cross-disciplinary mathematics based on holographic principle}, Morikita, 2016. [written in Japanese.]
\end{thebibliography}
\end{document}